\documentclass[preprint,showpacs,12pt,aps]{revtex4}
\usepackage{epsfig}
\usepackage{amssymb}
\usepackage{amsthm}

\begin{document}

\title{Effects of Dzyaloshinsky-Moriya Interaction on magnetism in Nanodisks}

\author{Zhaosen Liu$^{a}$\footnote{Email: liuzhsnj@yahoo.com},
Hou Ian$^b$ }

\affiliation{$^a$Department of Applied  Physics,
 Nanjing University of Information Science and Technology,   Nanjing 210044, China\\
$^b$ Institute of Applied Physics and Materials Engineering, FST,
University of Macau, Macau}

\begin{abstract}
We give a theoretical study on the magnetic properties of
monolayer nanodisks with both Heisenberg exchange and
Dzyaloshinsky-Moriya (DM) interactions. In particular, we
survey the magnetic effects caused by anisotropy, external
magnetic field, and disk size when DM interaction is present by
means of a new quantum simulation method  based on mean-field
theory. This computational approach finds that uniaxial anisotropy
and transverse magnetic field enhances the net magnetization as
well as increases the transition temperature of the vortex phase
while preserving the chiralities of the nanodisks. Whereas, when
the strength of DM interaction is sufficiently strong for a given
disk size, magnetic domains appear within the circularly bounded
region, which vanish and gives in to a single vortex when a
transverse magnetic field is applied. The latter confirms the
magnetic skyrmions induced by the magnetic field as observed in
the experiments.
\end{abstract}


\maketitle

\section{Introduction}
Dzyaloshinsky-Moriya (DM) interaction \cite{Dzy,Moriya}, which arises
from the spin-orbit scattering of electrons in an inversely asymmetric
crystal field, exists in systems with broken inversion symmetry, such
as the metallic alloys with B20 structure \cite{Tonomura,Yu,Wilhelm,Yu2,Muhlbauer}and
the surface or interface of magnetic multi-layers \cite{Heinze,Bode}.
This interaction is able to induce chiral spin structures, such as
skyrmions \cite{Skyrme}, and other exotic properties, providing the
basis for developing new spintronic devices. For instance, in the
confined structures such as magnetic nanodisks and nanostripes, DM
interaction can produce stable out-of-plane magnetizations as well
as in-plane left-handed and right-handed magnetic vortices \cite{Wachowiak,Shinjo,Choe,Im}.

These exotic magnetic structures have recently been experimentally
observed. For examples, Heinze et al. observed a spontaneous atomic-scale
magnetic ground-state skyrmion lattice in a monolayer Fe film at a
temperature about 11 K \cite{Heinze}; Yu et al.~have obtained a
skyrmion crystal near room-temperature in FeGe by applying a strong
magnetic field \cite{Yu}. The latter shows the low-temperature constrain
\cite{Kiselev} which presents for most skyrmions in helimagnets produced
by external magnetic fields \cite{Munzer,Yu2,Tonomura} can be overcome.

In the present work, we use a self-consistent approach (SCA) developed
in recent years \cite{liujpcm,liuphye141,liuphye142,liupssb} to investigate
the special magnetic properties of monolayer nanodisks induced by
DM interactions, where the co-existence of Heisenberg exchange and
uniaxial anisotopy are assumed. The effectiveness of the approach
has been verified on a nanowire consisting of ferromagnetically coupled
3D ions \cite{liuphye141} and a DyNi$_{2}$B$_{2}$C nanoball \cite{liupssb}.
The algorithm for this approach, which we brief in Sec.~II, is based
on the principle of the least free energy. That is, while the algorithm
approaches the eventual equilibrium state, the magnetic moments of
the system are rotated and their magnitudes adjusted by the local
effective field to minimize the system energy spontaneously. 

Here, simulations using the SCA algorithm are performed on nanodisks
with different sizes and DM interaction strengths and under different
scenarios. We start with the simplest case in Sec.~III.A, where the
lattice is considered isotropic and study the effect of anisotropy
in Sec.~III.B. The case when both anistropy and external magnetic
field are taken in consideration is discussed in Sec.~III.C. We find
that a single right-handed (left-handed) magnetic vortex is generated
in the disk plane by a $D>0$ ($D<0$) DM interaction strength when
the disk size is sufficient small (10 times the lattice constant).
In addition, an external magnetic field normal to the disk plane was
able to preserve the vortex structure above the transition temperature
at $T_{M}\approx$ 2.92 K, making the vortex still observable at 5
K.

The considerations above assume a relatively weak DM interaction where
the chiralities of eddy structures are preserved. In Sec.~III.E and
Sec.~III.F, we further increase the interaction strength while simultaneously
increase the disk size to a few times the DM length in diameter. We
find that multiple magnetic vortices, separated by strips, were formed.
The sizes of these magnetic domains agree well with the present theory.
In addition, the effect of DM interaction can be cancelled by an external
magnetic field normal to the disk, letting the multiple domains merge
to a single in-plane vortex. This finding concides with the experimental
observations \cite{Munzer,Yu,Yu2,Tonomura}.

\section{Self-Consistent Approach}

 The   nanosystems  we investigated in this work can be described
   with  a Hamiltonian \cite{YMLuo}
\begin{equation}
{\cal H} = -\frac{1}{2}\left[\sum_{i,j\neq i}{\cal J}_{ij}{
\vec{S}_i \cdot }\vec{S}_j - \sum_{i,j\ne i
}D_{ij}{\vec{r}_{ij}\cdot(\vec{S}_i\times}\vec{S}_j)\right]-K_A\sum_i(
{\vec S}_i\cdot \hat{n})^2-\mu_Bg_S{\vec B }\cdot\sum_i{\vec
S}_i\;, \label{hamil}
\end{equation}
where the first and second terms represent the Heisenberg exchange
and DM interactions with strength of ${\cal J}_{ij}$ and $D_{ij}$
between each pair of neighboring spins situated at the $i-$th and
$j-$th site, respectively, the third term denotes the uniaxial
anisotropy along the  $\hat{n}$ direction,  and the last term  is
the Zeeman energy of the system within   the applied magnetic
field ${\vec B}$. For simplicity, we consider here a few   round
monolayer   nanodisks consisting of $S$ = 1 spins, which interact
only with their nearest neighbors with   uniform strength, that
is, ${\cal J}_{ij}$ = ${\cal J}$ and $D_{ij} = D$ over each disk
which is assumed  to be in  the $xy$-plane, and the anisotropy is
perpendicular to the disk   plane, i.~e.  along the $z$-axis. In
the above Hamiltonian, the spins are quantum operators instead of
the classical vectors. Since $S = 1$, the matrices of the three
spin components are given by
\begin{eqnarray}
S_x =\frac{1}{2}\left(
\begin{array}{ccc}
0 &  \sqrt{2}  & 0\\
\sqrt{2} & 0 & \sqrt{2}  \\
0 & \sqrt{2}  & 0\\
\end{array}
\right)\;, & S_y =\frac{1}{2i}\left(
\begin{array}{ccc}
0 &  \sqrt{2}  & 0\\
-\sqrt{2} & 0 & -\sqrt{2}  \\
0 & \sqrt{2}  & 0\\
\end{array}
\right)\;, & S_z = \left(
\begin{array}{ccc}
1 &  0  & 0\\
0 & 0 & 0 \\
0 & 0  & -1\\
\end{array}
\right)\;,
\end{eqnarray}
respectively. And according to quantum theory,    the thermal
average of any physical quantity $A$ must be evaluated with
\begin{equation}
\langle A\rangle = \frac{\sum_n\langle \varphi_n
|\hat{A}\exp(-\varepsilon_n/k_BT)|\varphi_n\rangle}{\sum_n\exp(-\varepsilon_n/k_BT)}\;,
\label{avq}
\end{equation}
where    $\varepsilon_n$ and  $\varphi_n$ are  the eigenenergy and
eigenfunction,   respectively, of the related  Hamiltonian, for
instance, of the considered spin.

All of our recent simulations employing the SCA approach    were
started from a random magnetic configuration  and  a temperature
above the magnetic transition, then carried out stepwise down to
very low temperatures with a   temperature step $\Delta T < 0$.
This point  is {\it crucial}, since we believe that at high
temperatures the interactions among the spins with small
magnitudes are considerably weak,  the thermal interaction is
strong enough to help the spins overcome the energy barriers, so
that the code can avoid being trapped in local energy minima, and
finally converge down to the equilibrium state of the system with
globally least total (free) energy. Obviously, to ensure
 thermal energy of the spins sufficiently strong at all subsequent
temperatures, $|\Delta T|$ cannot be too large. At any temperature
$T$, if the difference $|\langle \vec{S}'_i\rangle - \langle
\vec{S}_i\rangle|$ between two successive iterations for every
spin is less than a very small given value $\tau_0$, convergency
is considered to be reached.

\section{Calculated Results}

\subsection{Magnetic Properties of a Nanodisk Without Anisotropy}
In order to  visualize  the calculated spin configurations
clearly, we   first conducted simulations  with  the SCA approach
for a very tiny round monolayer nanodisk, whose   radius $R$ =
10$a$, where $a$ is the side length of the square crystal unit
cell. Other parameters were assigned to ${\cal J}/k_B = 1$ K,
$D/k_B = \pm 0.1$ K, and $\Delta T$ = -0.02 K, respectively.

For comparison,    the influences of both uniaxial anisotropy and
the external magnetic field was neglected in simulations at the
beginning. Figure 1 displays our calculated spontaneous thermally
averaged $\langle S_z\rangle $, $\langle S_{x}\rangle $ and
$\langle S_{y}\rangle $ for the round nanodisk with the parameters
given below  the figures. The on-plane DM interaction has indeed
induced out-plane magnetic moments \cite{Wachowiak,Shinjo}, which
are stronger than the on-plane components in the magnetic phase as
displayed in the two sub-figures. As temperature rises, the three
components decay monotonously until the transition temperature
$T_M \approx $ 2.66 K, and   the saturated value of $\langle
S_z\rangle$ at very low temperatures is much less than the maximum
value $S$ =  1.

\begin{figure*}[htb]
 \centerline{\epsfig{file=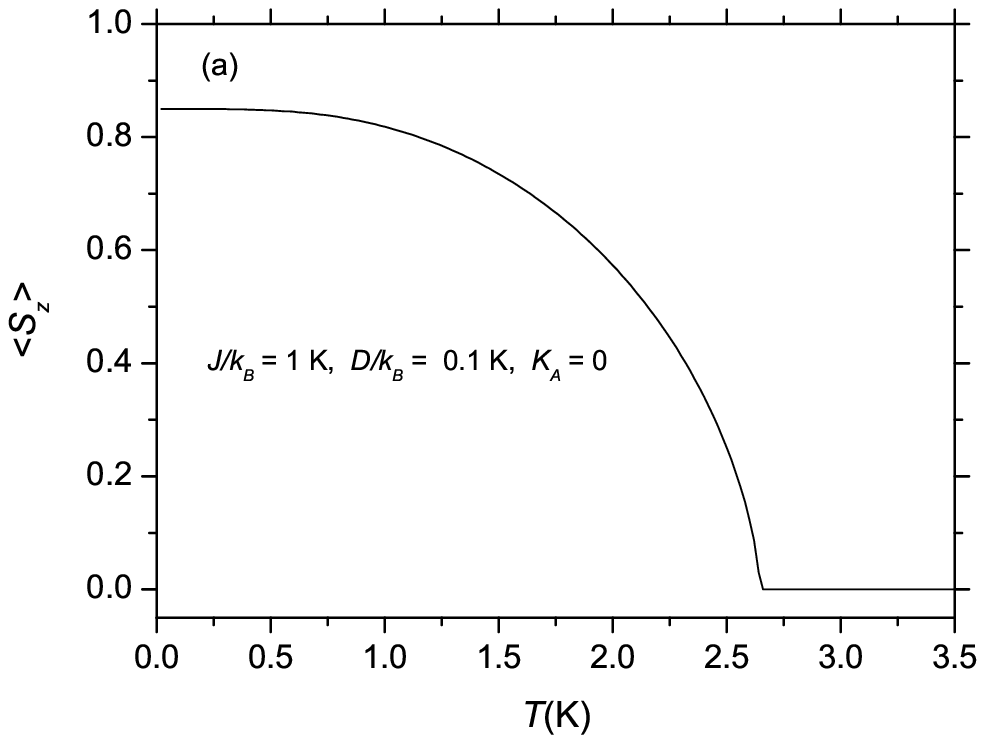,width=.45\textwidth,height=6.5cm,clip=}
 \epsfig{file=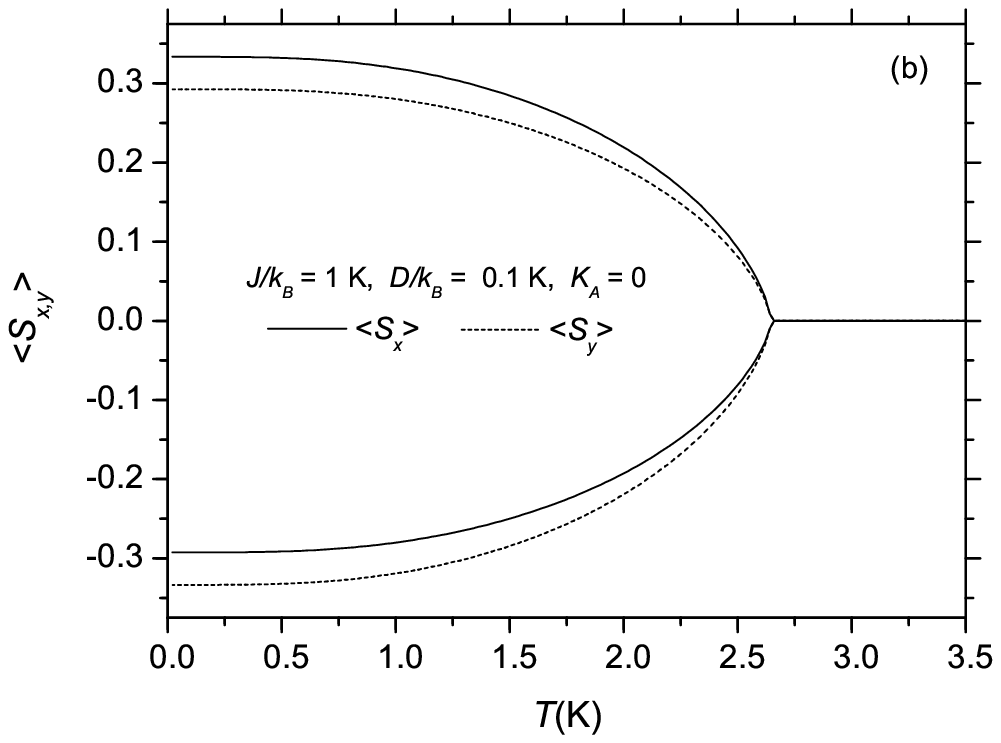,width=.45\textwidth,height=6.5cm,clip=}}
\begin{center}
\parbox{14cm}{\small{{\bf Figure 1.}  Calculated spontaneous
(a)   $\langle S_z\rangle $,   (b) $\langle S_{x}\rangle $ and
$\langle S_{y}\rangle $ for the monolayer nanodisk. The used
parameters are $R$ = 10a, ${\cal J}/k_B$ = 1 K, and $D/k_B$ = 0.1
K, respectively.}}
\end{center}
\end{figure*}

\begin{figure*}[htb]
\centerline{
 \epsfig{file=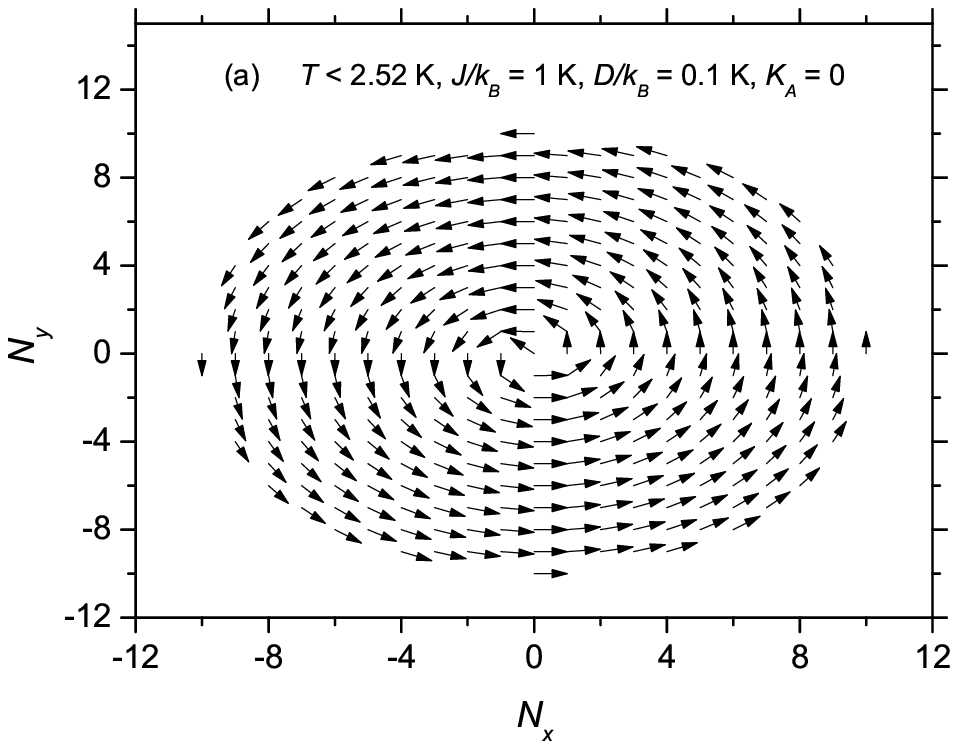,width=.45\textwidth,height=7cm,clip=}
\epsfig{file=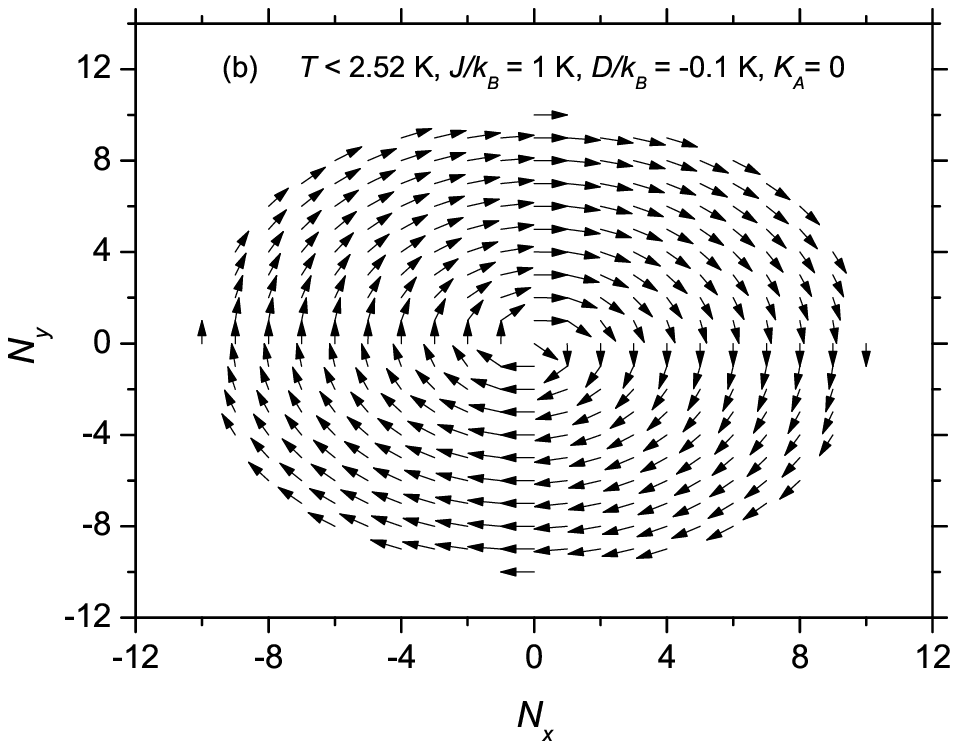,width=.45\textwidth,height=7cm,clip=} }
\begin{center}
 \parbox{16cm}
{\small{{\bf Figure 2.} Calculated spin configurations projected
onto the  $xy$-plane as (a) $D/k_B$ = 0.1 K, and (b) $D/k_B$ =
-0.1 K in the absence of external magnetic field. Other used
parameters are $R$ = 10$a$, ${\cal J}/k_B$ = 1 K, $K_A/k_B$ = 0 K,
respectively. }}
\end{center}
\end{figure*}

Due to the DM interaction, magnetic vortices are formed on the
nanodisks as depicted in Figure 2, they were obtained  with
$D/k_B$ assigned to  $\pm$ 0.1 K  respectively, but   other
parameters remained the same as before. In the case of $D/k_B =
0.1 $ K, the directions of $\langle S_z\rangle$ and the magnetic
vortex in the $xy$-plane meet the right-hand spiral rule. But when
$D/k_B = -0.1 $ K,  the chirality of the vortex, as displayed in
Figure 2(b), is reversed  \cite{Bode}. However, the calculated
$\langle S_z\rangle$ curve is now still exactly identical to that
obtained with $D/k_B = 0.1 $  as plotted in Figure 1(a).

\subsection{Effects of Uniaxial Anisotropy}

By taking  account of  a strong uniaxial anisotropy of $K_A/k_B$
=1 K normal to the disk plane, we performed further simulations.
As displayed in Figure 3(a),   the  transition temperature is now
increased to $T_M \approx $ 2.92 K, and $\langle S_z\rangle $
becomes almost saturated at very low temperatures. Therefore,
$\langle S_z\rangle $ has been considerably enhanced by the
uniaxial anisotropy along the same direction. But for the same
sake, as shown in Figure 3(b), $\langle S_{x}\rangle $ and
$\langle S_{y}\rangle $ have been
\begin{figure*}[htb]
 \centerline{\epsfig{file=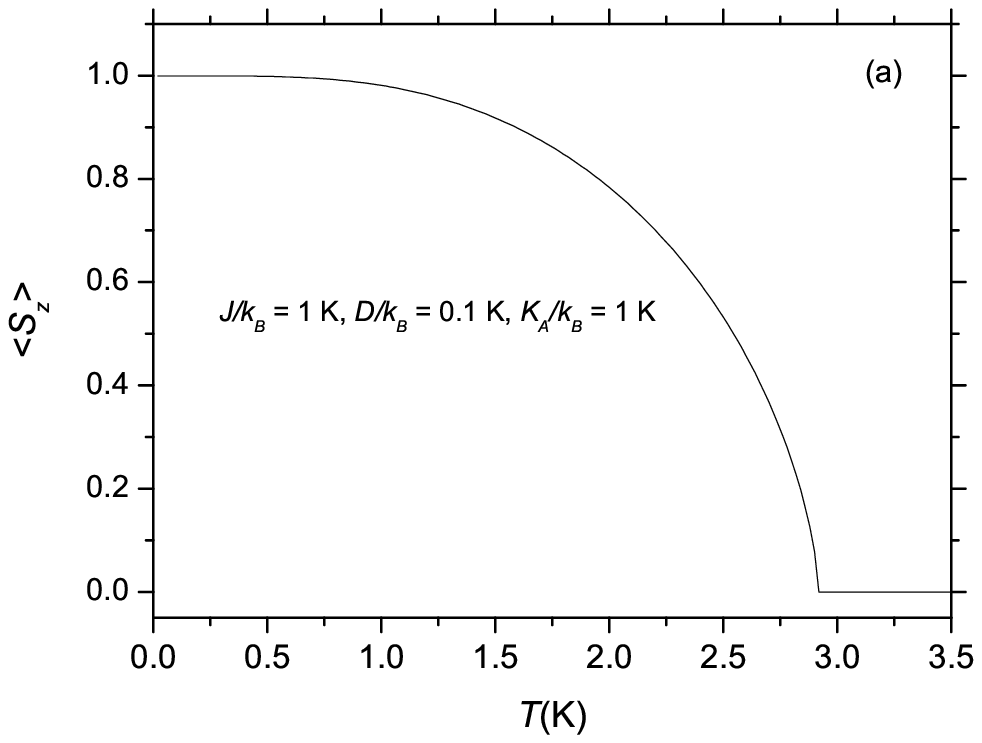,width=.45\textwidth,height=7cm,clip=}
 \epsfig{file=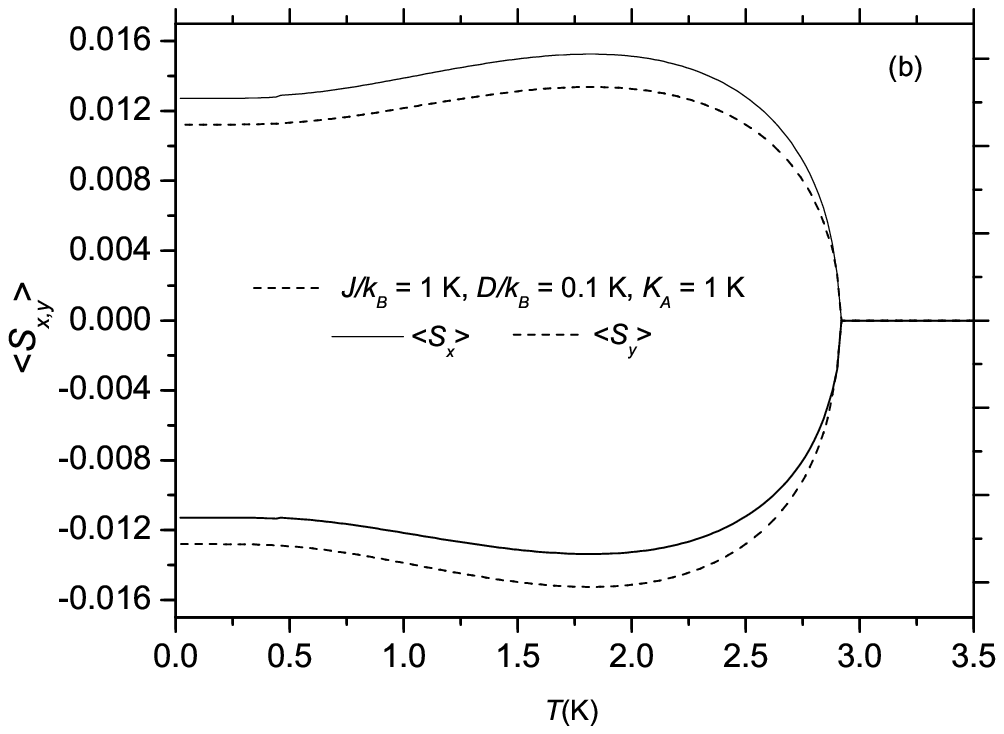,width=.45\textwidth,height=7cm,clip=}}
\begin{center}
\parbox{16cm}{\small{{\bf Figure 3.}
 Calculated spontaneous
(a)   $\langle S_z\rangle $, (b) $\langle S_{x}\rangle $ and
$\langle S_{y}\rangle $ for the monolayer   nanodisk.
 Here $R$ = 10$a$, ${\cal J}/k_B$ = 1 K, $D/k_B$ = 0.1 K,  and  $K_A/k_B$ =1 K , respectively.
 }}
\end{center}
\end{figure*}
considerably reduced by the anisotropy, and their shapes   also
changed a lot. On the other hand, when $ D/k_B$ = -0.1 K but other
parameters are unchanged, we can still get very similar three
components: the $\langle S_z\rangle $ curve remains same, but
$\langle S_{x}\rangle $ and $\langle S_{y}\rangle $ curves just
exchange their positions.  In this case, we also obtained  almost
same magnetic structures as those shown in Figure 2. To short the
text, these results are not plotted here.  Once again, when $D/k_B
= 0.1 $ K, the directions of $\langle S_z\rangle$ and the magnetic
vortex satisfies the right-hand screw rule; but while $D/k_B =
-0.1 $ K, the chirality of the vortex is reversed \cite{Bode}.
 As $T$ = 0.28 K we observed  in the both cases that the vectors at the   lattice
center disappeared. This   fact means  that, at very low
temperatures, the spin at the center has been rotated by the
uniaxial anisotropy completely to the direction normal to the
disk.

\begin{figure*}[htb]
\centerline{
\epsfig{file=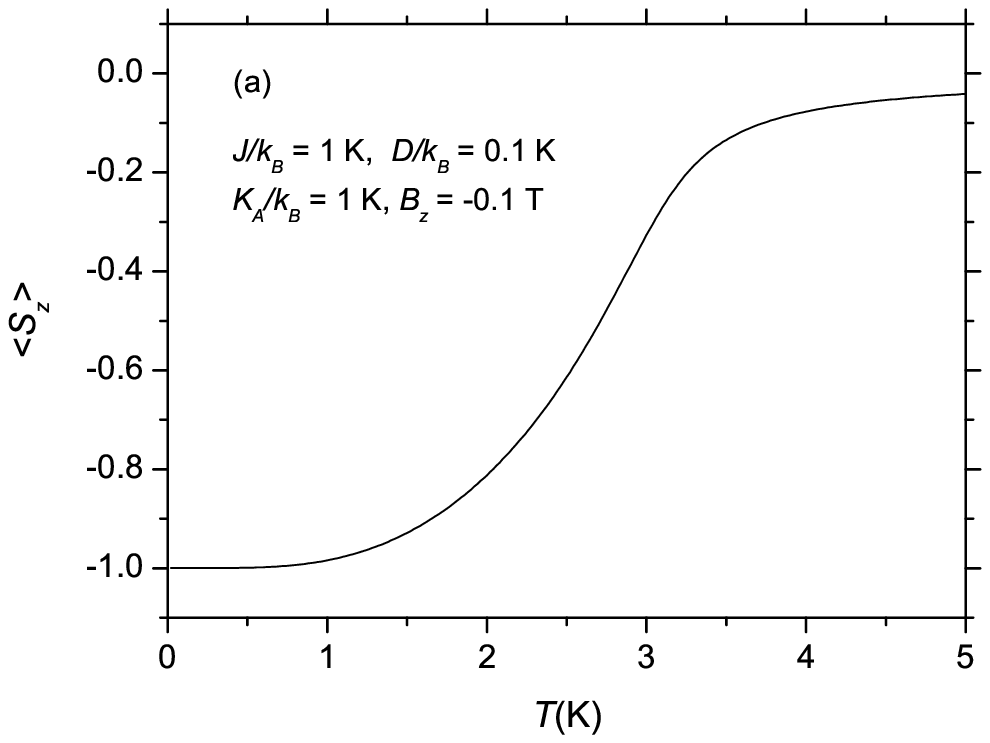,width=.45\textwidth,height=7cm,clip=}
\epsfig{file=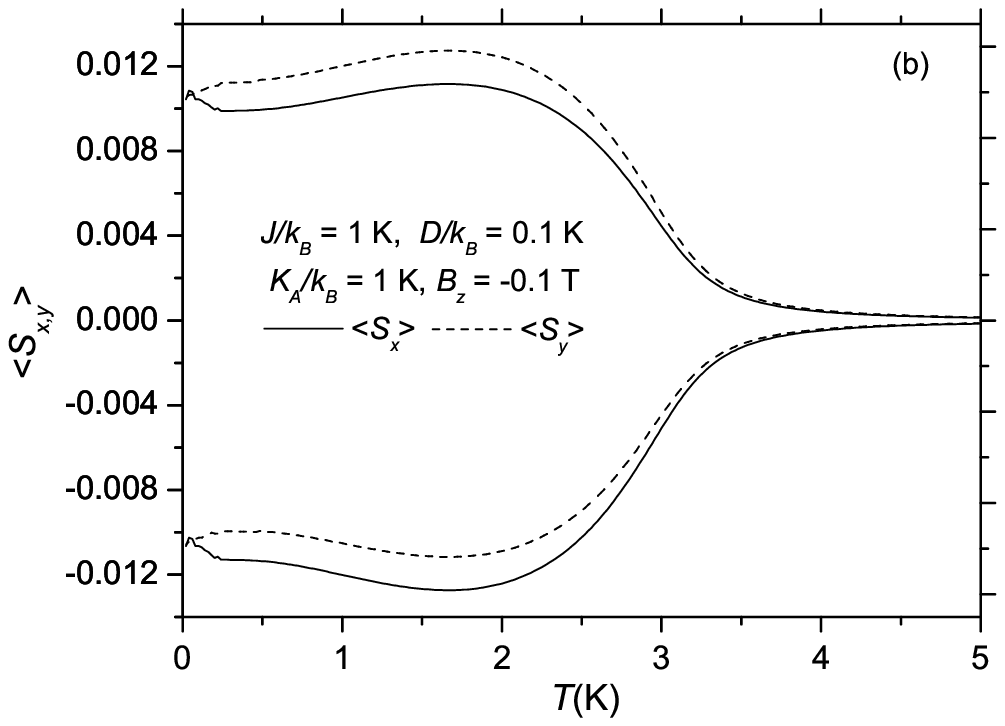,width=.45\textwidth,height=7cm,clip=}}
\begin{center}
\parbox{16cm}{\small{{\bf Figure 4.} Calculated
(a)   $\langle S_z\rangle $,   (b) $\langle S_{x}\rangle $ and
$\langle S_{y}\rangle $ for the monolayer   nanodisk  in a
magnetic field  of 0.1 Tesla applied antiparallel to the $z$ axis.
Other parameters are $R$ = 10$a$,   $D/k_B$ = 0.1 K, ${\cal
J}/k_B$ = 1 K, and $K_A/k_B$ = 1 K, respectively. }}
\end{center}
\end{figure*}

\subsection{Effects of External Magnetic Field}

Later on, a magnetic field  antiparallel to the direction of the
spontaneous $\langle S_z\rangle$ was considered to do simulations.
As displayed in Figure  4.
\begin{figure*}[htb]
\centerline{
\epsfig{file=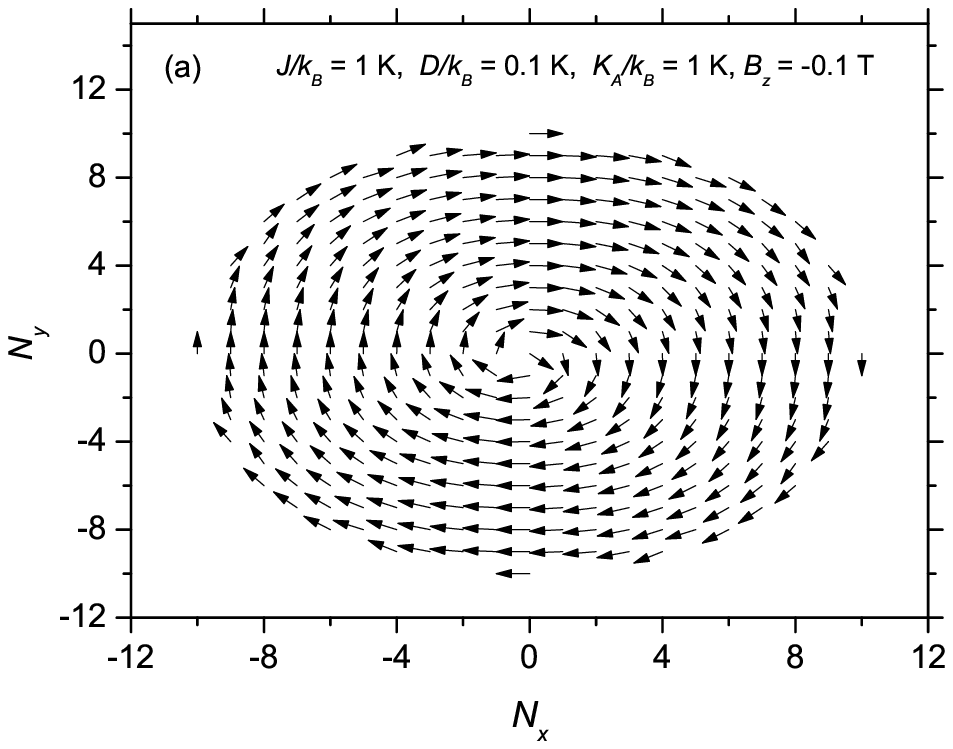,width=0.45\textwidth,height=7cm,clip=}
\epsfig{file=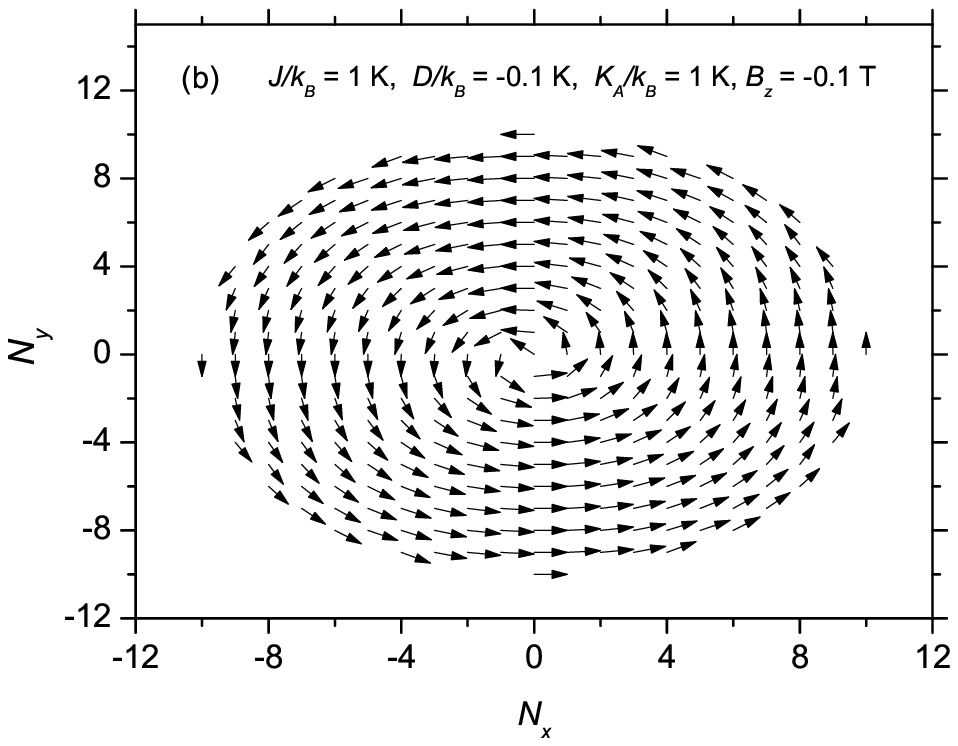,width=0.45\textwidth,height=7cm,clip=} }
\begin{center}
\parbox{16cm}{\small{{\bf Figure 5.}
Calculated spin  configurations projected onto the $xy$-plane
within an external magnetic field of 0.1 Tesla applied
antiparallel to the $z$ direction, as (a) $D/k_B$ = 0.1 K, and (b)
$D/k_B$ = -0.1 K. Other used parameters are $R$ = 10$a$, ${\cal
J}/k_B$ = 1 K, and $K_A/k_B$ = 1 K,  respectively.  }}
\end{center}
\end{figure*}
 the spontaneous $\langle S_z\rangle$ has been rotated by   the applied magnetic field for 180$^0$
 to its direction. And  no matter $D/k_B$ = 0.1 or -0.1 K, we
 obtained   the   exactly  same $\langle S_{z}\rangle$,
$\langle S_{x}\rangle$ and $\langle S_{y}\rangle$ curves, the only
difference was that  as $D$ changed the sign  the two transversal
components   exchanged their positions. The external magnetic
field has considerably modified  the shapes of the three curves
especially in the region around the transition temperature, and
strongly enhanced   the $z$-components of the magnitization, so
that it  persists  to  much higher  temperatures. As depicted in
Figure 5, the direction of the magnetic vortex in the $xy$-plane,
which is stable below $T$ = 4  K and can still  be observed at $T$
= 5 K, has also been reversed by the applied magnetic field,
however their chiralities are well conserved.

\begin{figure*}[htb]
\centerline{
 \epsfig{file=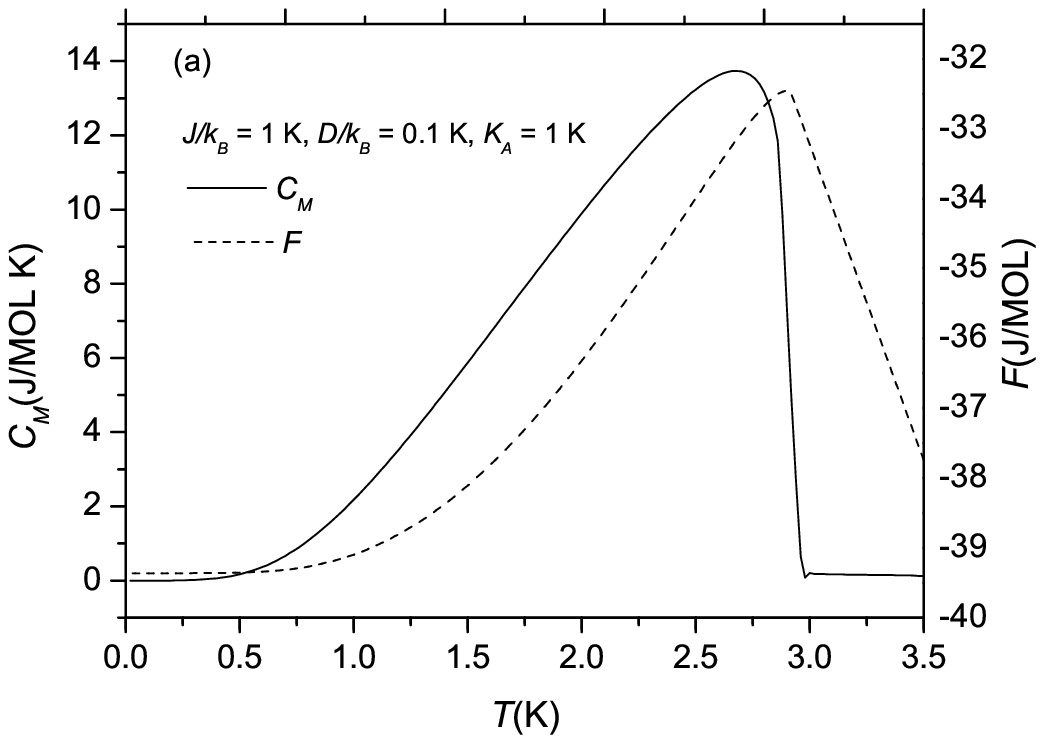,width=.45\textwidth,height=7cm,clip=}
\epsfig{file=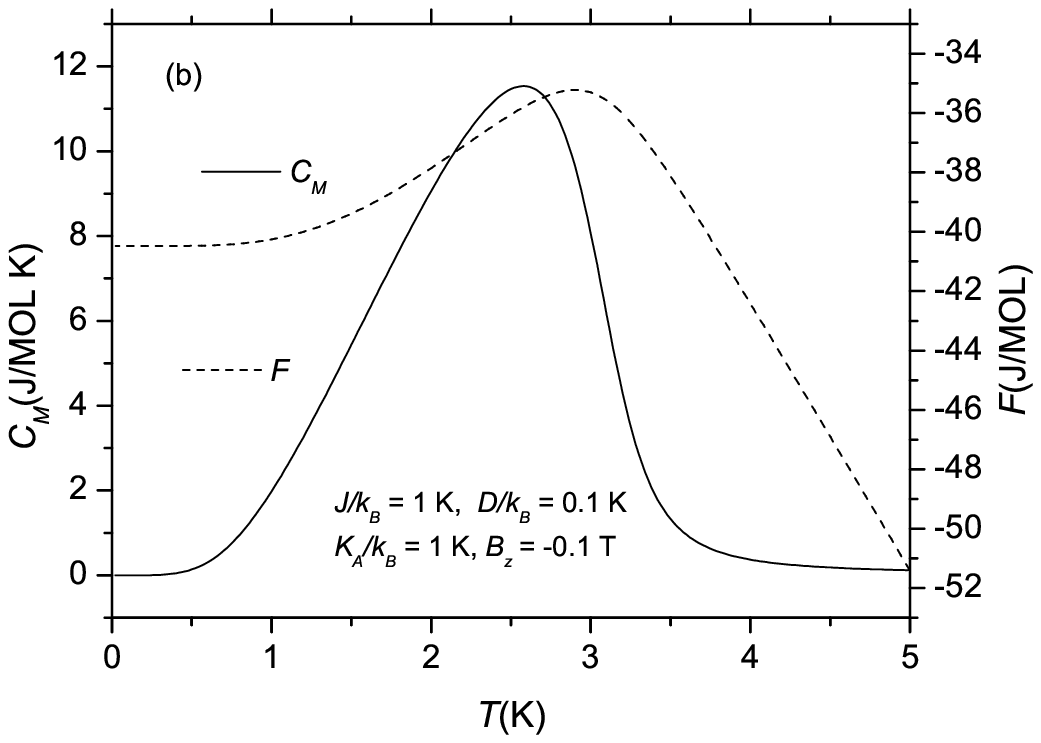,width=.45\textwidth,height=7cm,clip=} }
\begin{center}
 \parbox{16cm}
 {\small{{\bf Figure 6.}  Calculated specific heats and free energies per mole of spins
 when (a) no external magnetic is applied,  and  (b)   an external magnetic field  of 0.1
 Tesla  is   applied antiparallel to the  $z$ axis.
 Other used parameters are  $R$ = 10$a$, ${\cal J}/k_B$ = 1 K,  and $K_A/k_B$ = 1 K, respectively.}}
 \end{center}
\end{figure*}

\subsection{Thermodynamic Properties of the Nanodisk}

  The total free energy $F$, total energy
$E$, entropy $S_M$ and specific heat $C_M$ of  these  canonical
systems can be evaluated with
\begin{eqnarray}
 F = -k_BT\log Z_N,\space\space & E = -\frac{\partial}{\partial\beta}\log Z_N\;,\nonumber\\
 S_M = \frac{E}{T}+ k_B\log Z_N, &
 C_M= T\left(\frac{\partial S_M}{\partial T}\right)_B\;,
\end{eqnarray}
successively,  where $\beta =1/(k_BT)$ and $Z_N$ is the partition
function of the whole system.  The first three quantities can be
calculated during the simulation, but $C_M$ must be  computed by
using the last formula after the simulation for the whole
temperature range has been completed.  To shorten the text, only
the free energies and specific heats  calculated with the SCA
approach, both in the absence and presence of external magnetic
field exerted antiparallel to the spontaneous $\langle
S_z\rangle$, are plotted in Figure 6 for comparison. The
parameters used   in the simulation are given below the figure.
The magnetic behaviors of the nanodisk have been considerably
modified by the applied magnetic field, so that the curves near
the transition temperature  changes smoothly as displayed in
Figure 6(b).  The   peak  in each $C_M$ curve  is the evidence of
phase transition  within the narrow temperature interval.

\subsection{Effects of    DM Interaction Strength and   External Magnetic Field}
To deal with the systems with magnetically eddy structures, DM
length has been introduced and defined as $\zeta = {\cal J}/D$,
which is related to the size of self-organized structures
\cite{Kwon13}. So, as the size of the disk is lager than $\zeta$
in the unit of lattice parameter $a$, more magnetic domains, such
as strips and vortices, will be formed.

By increasing the DM Interaction to   $D/k_B$ = 0.3 K and assuming
that no external magnetic field is applied,   simulations were
then done for the nanodisk.   The calculated $\langle S_z\rangle
$, $\langle S_{x}\rangle $ and $\langle S_{y}\rangle $ are plotted
in Figure 7. Now $\zeta = {\cal J}/D \approx 3.333 $, and $R >
\zeta$, so it is expected more self-organized structures will
appear.     The sudden changes of $\langle S_z\rangle$ and
$\langle S_z\rangle$ around 1.3 K indicate that a phase transition
occurs nearby. Indeed, above 1.3 K, the magnetic vortex, as shown
in Figure 8(a), can still be formed around the disk center.
However, below that temperature, a magnetic strip appears between
the two magnetic vortices, as displayed in Figure 8(b), whose
upper and lower parts are excellently symmetric about the
horizontal line $N_y$ = 0.

\begin{figure*}[htb]
\centerline{
 \epsfig{file=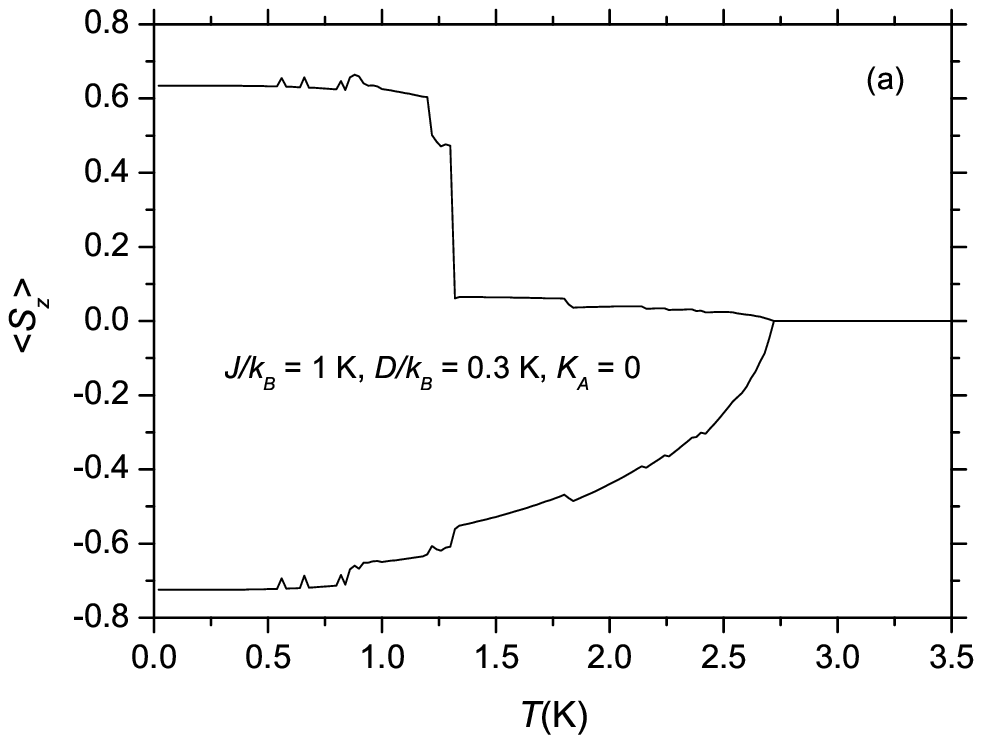,width=.45\textwidth,height=7cm,clip=}
\epsfig{file=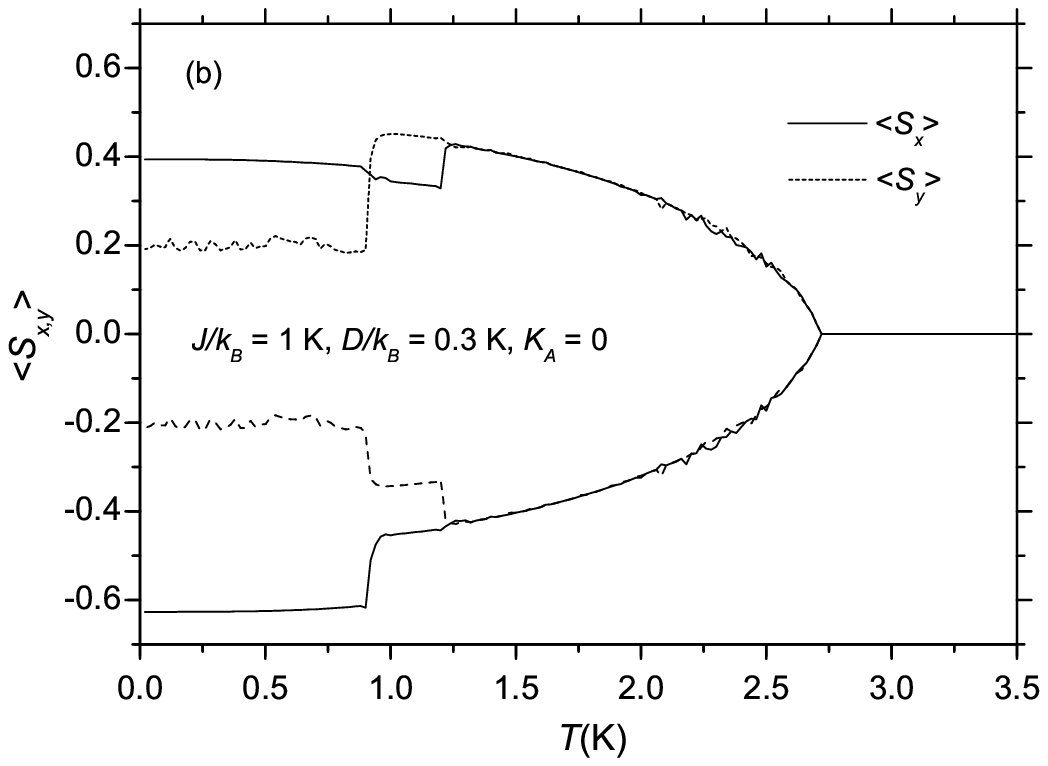,width=.45\textwidth,height=7cm,clip=} }
\begin{center}
 \parbox{16cm}
 {\small{{\bf Figure 7.}  Calculated spontaneous
(a)   $\langle S_z\rangle $,   (b) $\langle S_{x}\rangle $ and
$\langle S_{y}\rangle $ for the monolayer   nanodisk with $D/k_B$
= 0.3 K. Other parameters are $R$ = 10$a$, ${\cal J}/k_B$ = 1 K,
and $K_A/k_B$ = 0 K.}}
 \end{center}
\end{figure*}

\begin{figure*}[htb]
\centerline{
 \epsfig{file=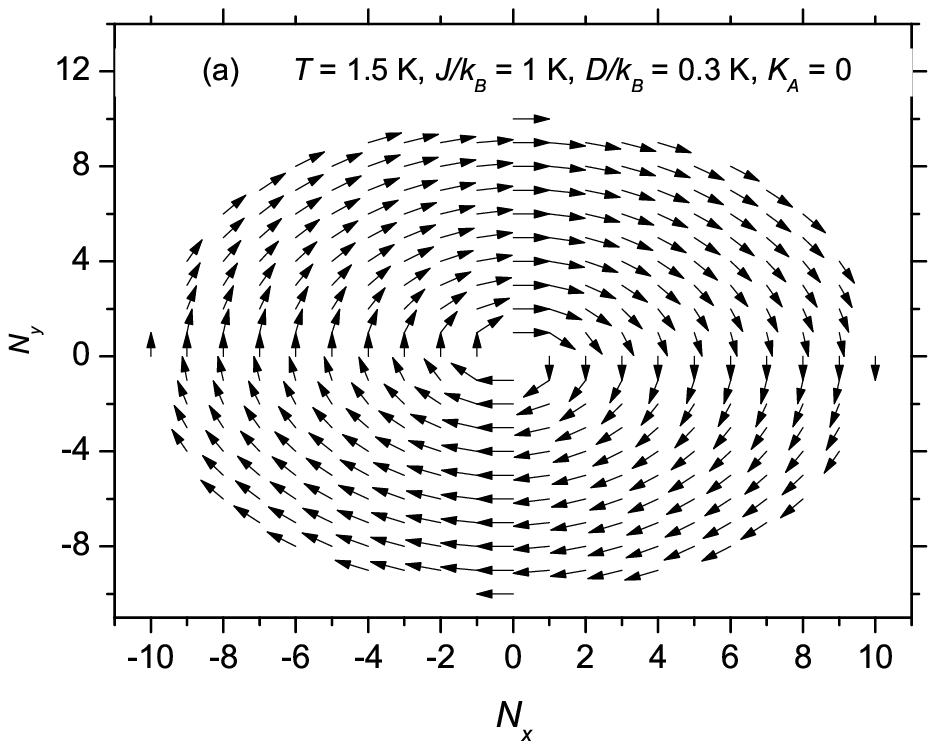,width=.45\textwidth,height=7cm,clip=}
\epsfig{file=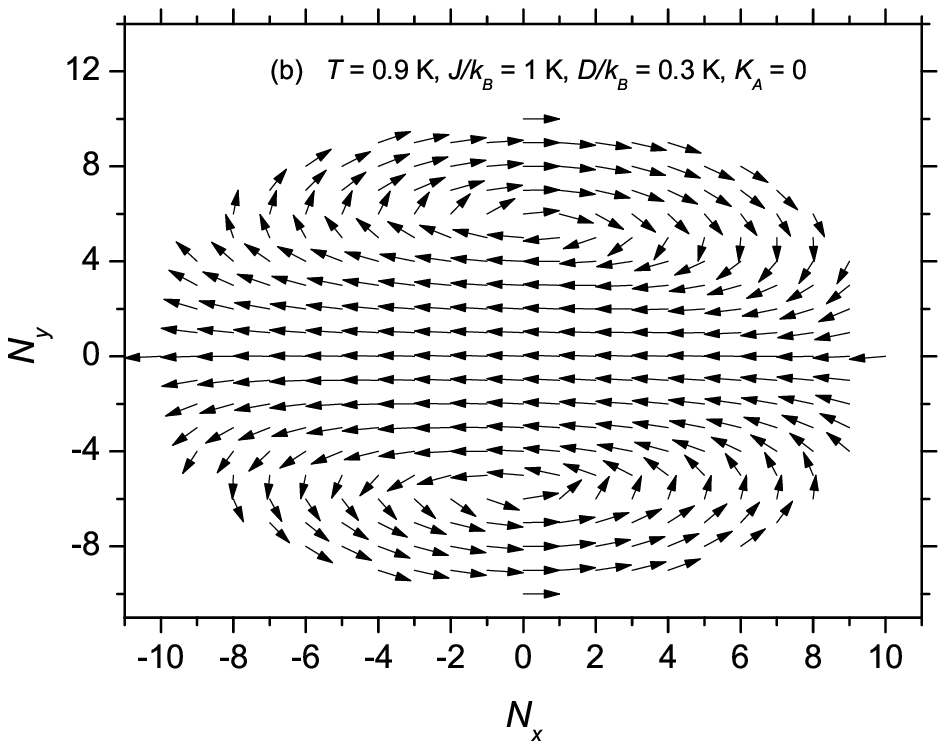,width=.45\textwidth,height=7cm,clip=}}
\begin{center}
 \parbox{16cm}
 {\small{{\bf Figure 8.} Calculated spontaneous spin configurations projected
onto the  $xy$-plane at (a) $T$ = 1.5 K, and (b) $T$ = 0.9 K,
respectively. Other used parameters are $R$ = 10$a$, ${\cal
J}/k_B$ = 1 K, $D/k_B$ = 0.3 K, and $K_A$ = 0 K. }}
 \end{center}
\end{figure*}
Therefore, as    DM interaction increases, the single vortex
structure becomes unstable, and magnetic domains will appear. It
was found in experiments that magnetic skyrmions could be induced
by external magnetic field  \cite{Munzer,Yu2, Tonomura,Yu}, which
means that the applied   field is able to stabilize the eddy
structure. To test this idea,   an external magnetic field of 0.1
tesla was considered to be applied in the $z$-direction to do
simulations for the nanodisk. Consequently, under the interaction
of the applied field, $\langle S_z\rangle $ is rotated to the
field direction, becoming positive in the whole temperature
region,   both $\langle S_z\rangle $ and $\langle S_y\rangle$ are
well symmetric about the $T$-axis, all  of them decay monotonously
with arising temperature as shown in Figure 9(a). These  magnetic
behaviors   indicate that a  single magnetic structure has been
generated on the disk by the applied magnetic field. Indeed, as
depicted in Figure 9(b),   we find now under the interaction of
the external magnetic field   a single magnetic vortex that
occupies the whole disk in the temperature range when $\langle
S_z\rangle $ is appreciable.

\begin{figure*}[htb]
\centerline{
 \epsfig{file=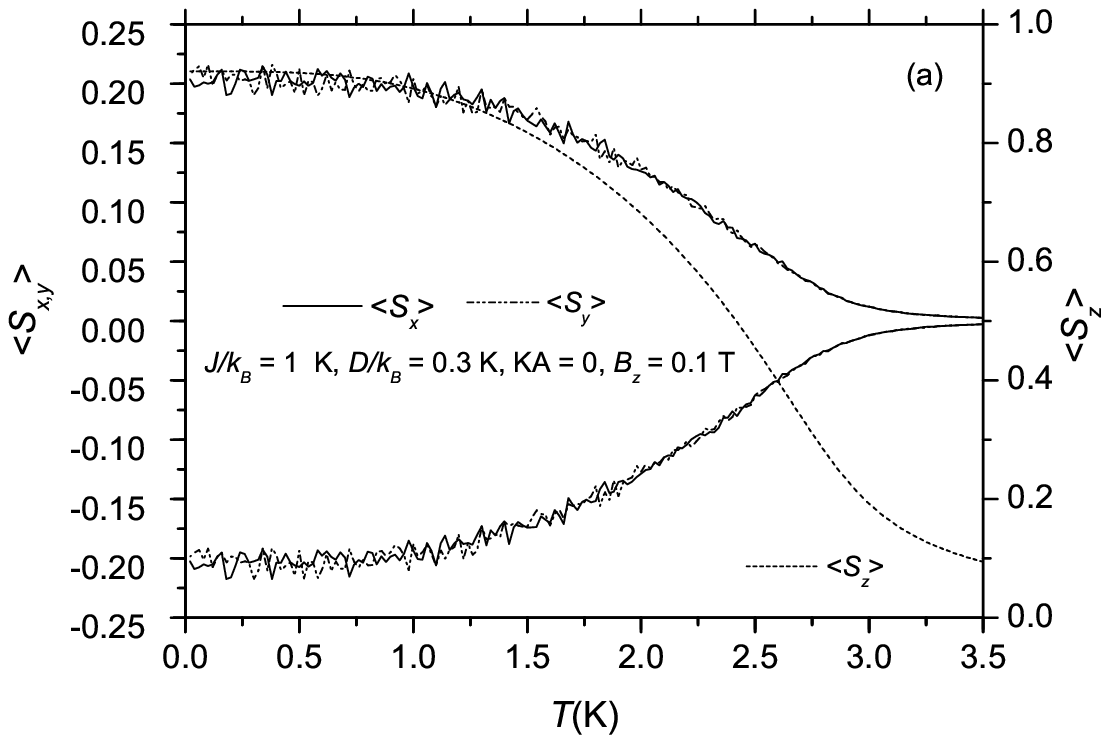,width=.45\textwidth,height=7cm,clip=}
\epsfig{file=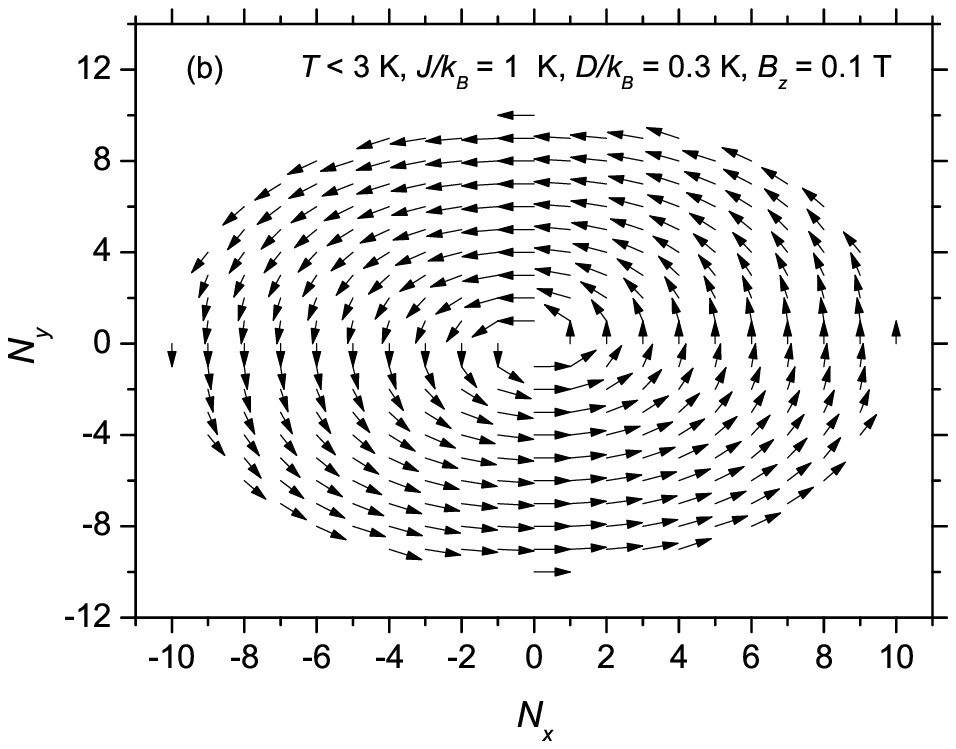,width=.45\textwidth,height=7cm,clip=} }
\begin{center}
 \parbox{16cm}
 {\small{{\bf Figure 9.}  Calculated
(a)      $\langle S_{x}\rangle $,  $\langle S_{y}\rangle $  and
$\langle S_z\rangle $, (b) the spin configuration for the
monolayer nanodisk in a magnetic field of 0.1 T applied along the
$z$ axis. Other parameters are $R$ = 10$a$, ${\cal J}/k_B$ = 1 K,
$D/k_B$ = 0.3 K, and $K_A$ = 0 K, respectively.}}
 \end{center}
\end{figure*}

\subsection{Effects of Disk Size and External Magnetic Field}

\begin{figure*}[htb]
\centerline{
 \epsfig{file=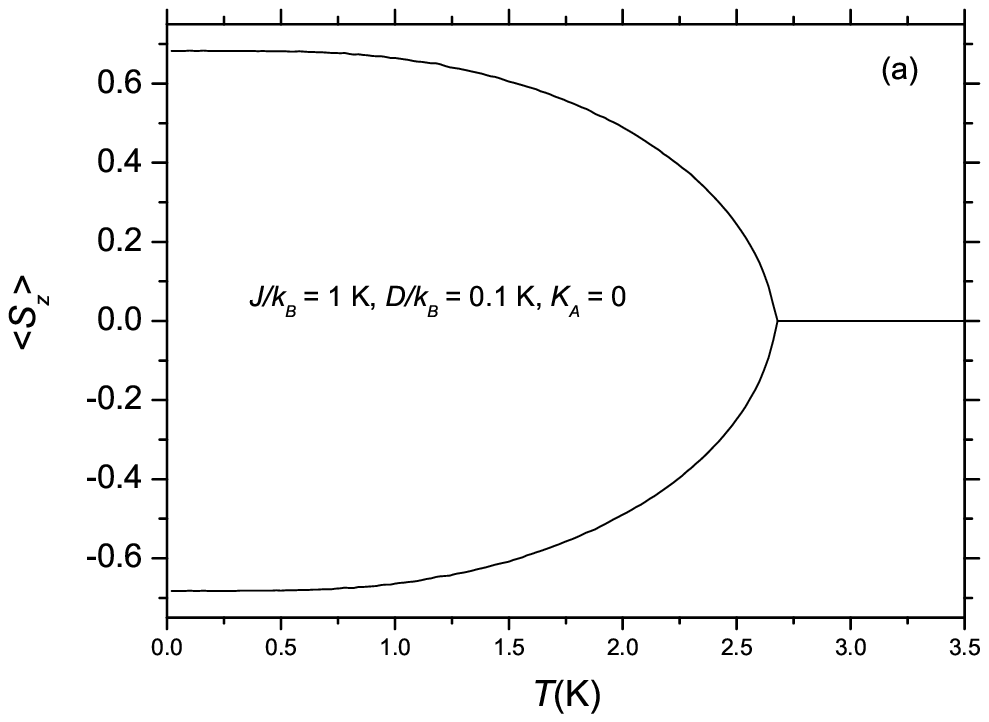,width=.45\textwidth,height=7cm,clip=}
\epsfig{file=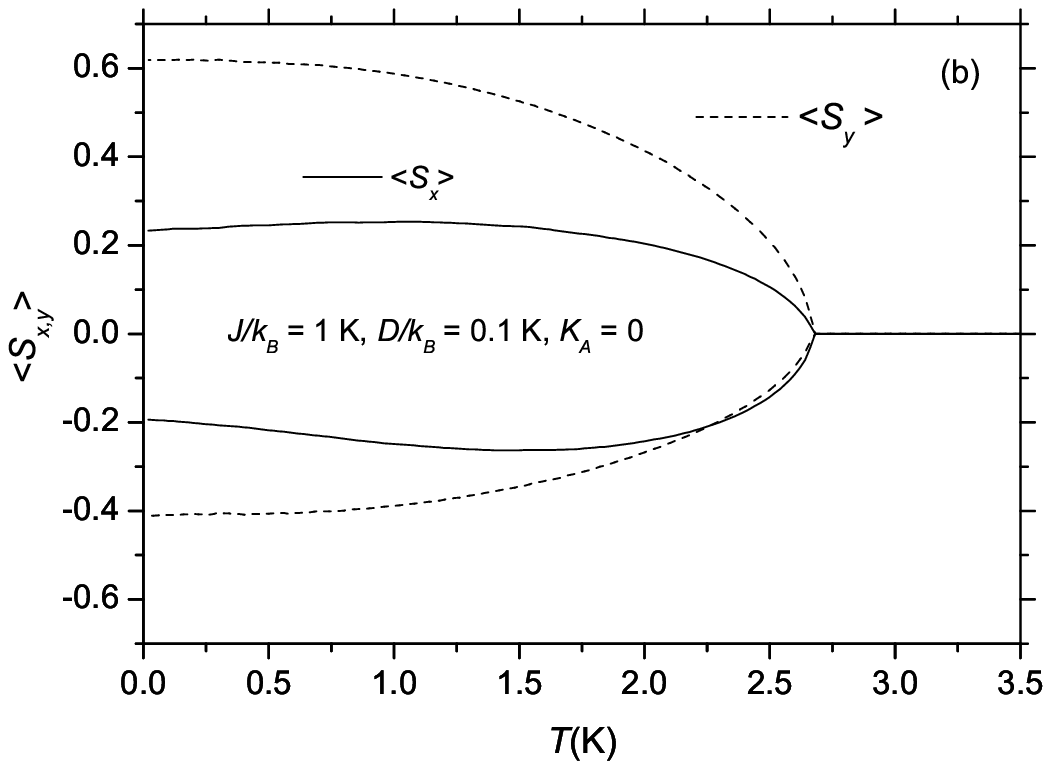,width=.45\textwidth,height=7cm,clip=} }
\begin{center}
 \parbox{16cm}
 {\small{{\bf Figure 10.}  Calculated spontaneous
(a)   $\langle S_z\rangle $,   (b) $\langle S_{x}\rangle $ and
$\langle S_{y}\rangle $ for a larger monolayer   nanodisk with $R$
= 30$a$. Other parameters are   $D/k_B$ = 0.1 K, ${\cal J}/k_B$ =
1 K, and $K_A$= 0.}}
 \end{center}
\end{figure*}

\begin{figure*}[htb]
\centerline{
 \epsfig{file=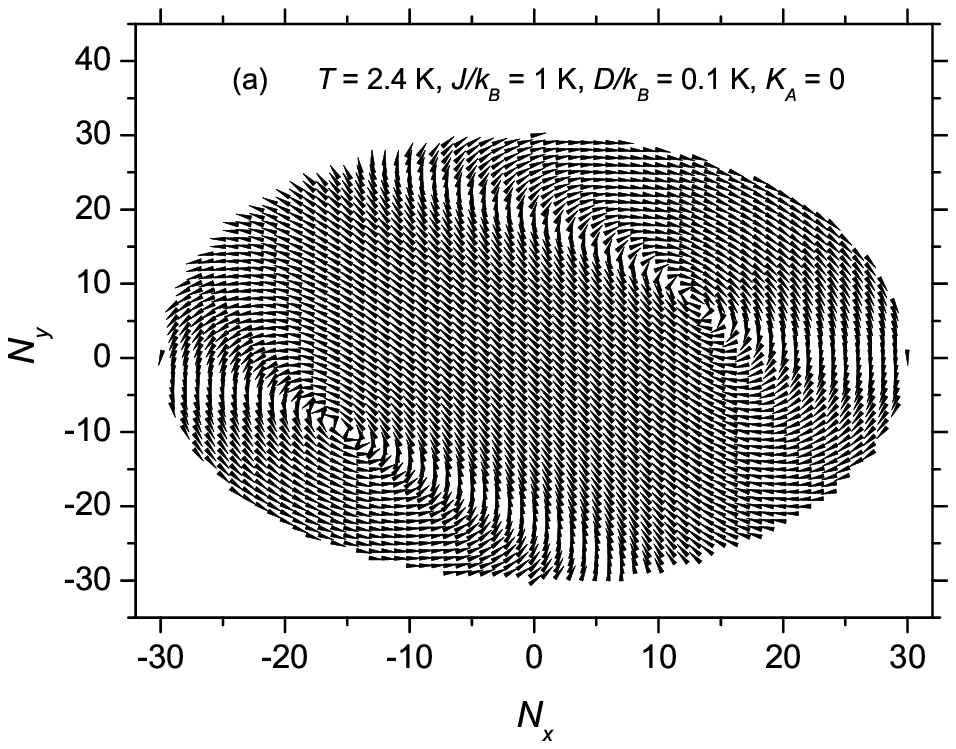,width=.7\textwidth,height=9cm,clip=}
 }
\centerline{
 \epsfig{file=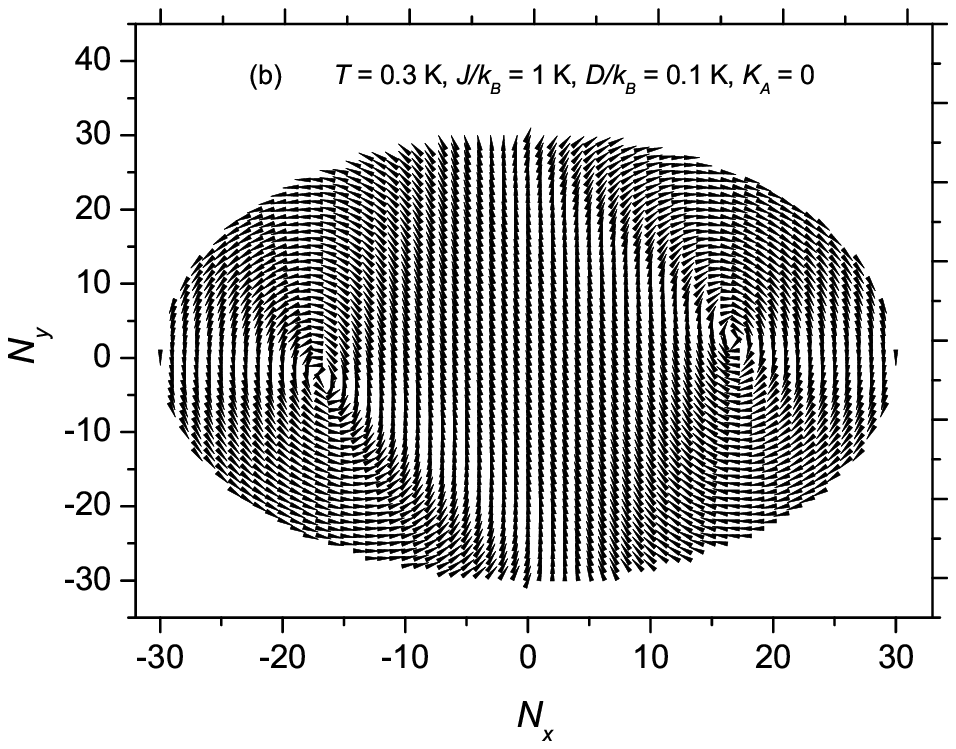,width=.7\textwidth,height=9cm,clip=}
 }
\begin{center}
 \parbox{16cm}
 {\small{{\bf Figure 11.} Calculated spontaneous spin  configurations projected onto the $xy$-plane
for the   nanodisk  with $R$ = 30$a$ at  (a) $T$ = 2.4 K, and  (b)
$T$ = 0.3 K, respectively.  Other used parameters are  ${\cal
J}/k_B$ = 1 K, $D/k_B$ = 0.1 K,  and $K_A$ = 0.} }
\end{center}
\end{figure*}
To further study the size effects, we  then considered a larger
monolayer nanodisk with $R$ = 30$a$,    ${\cal J}/k_B$ = 1 K,
$D/k_B$ = 0.1 K,  and $K_A$= 0, respectively,  and did simulations
by assuming  that no external magnetic was present firstly. In
this case, $\zeta = 10 < R$, so more magnetic structures will be
self-organized on the disk plane. Though $\langle S_z\rangle $ and
$\langle S_y\rangle $ decay monotonously with increasing
temperature, the presence of the both $\langle S_z\rangle \lessgtr
0 $ and especially the asymmetry of $\langle S_x\rangle $ and
$\langle S_y\rangle $, as shown in Figure 10, are the signs of
existence of more than one structures on the disk plane. Indeed,
as shown in Figure 11, two magnetic vortices are generated in our
simulation, they are separated by a magnetic strip. The two vortex
centers are rotated as temperature changes, their distance is
about 33$a$, which is about three times of $\zeta$, since three
structures, one strip and two vortices, are formed between the
centers.  The square lattice cell is four-fold rotationally
symmetric about the $z$-axis, but the asymmetric DM interaction
has reduced  the symmetry to two-fold, then only two vortices can
be observed on the disk plane, we guess.

\begin{figure*}[htb]
\centerline{
 \epsfig{file=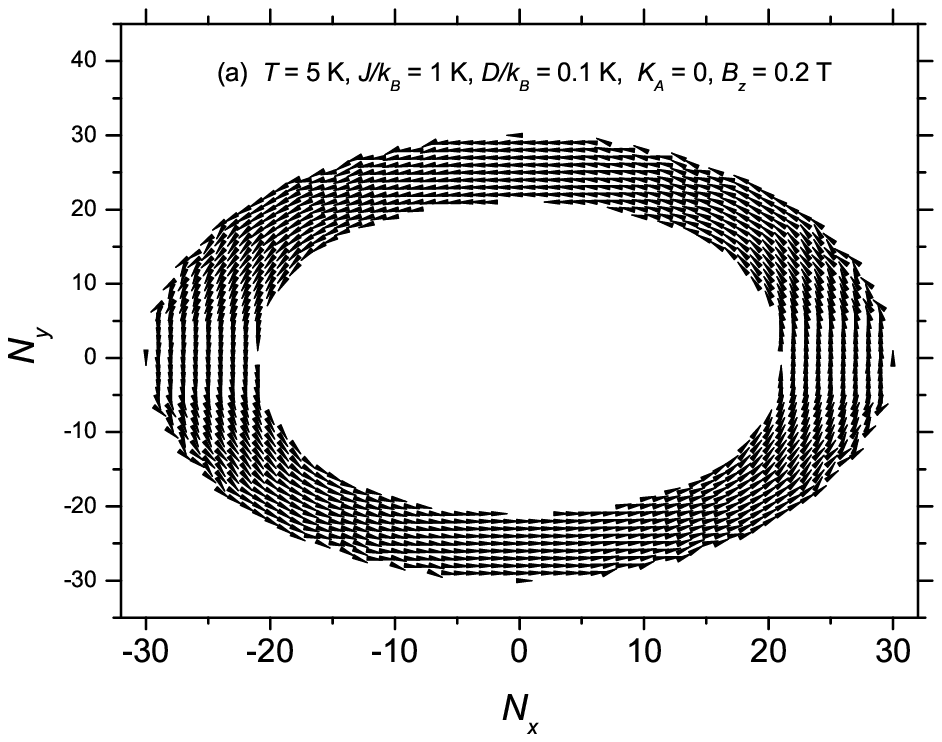,width=.7\textwidth,height=9cm,clip=}
 }
\centerline{
 \epsfig{file=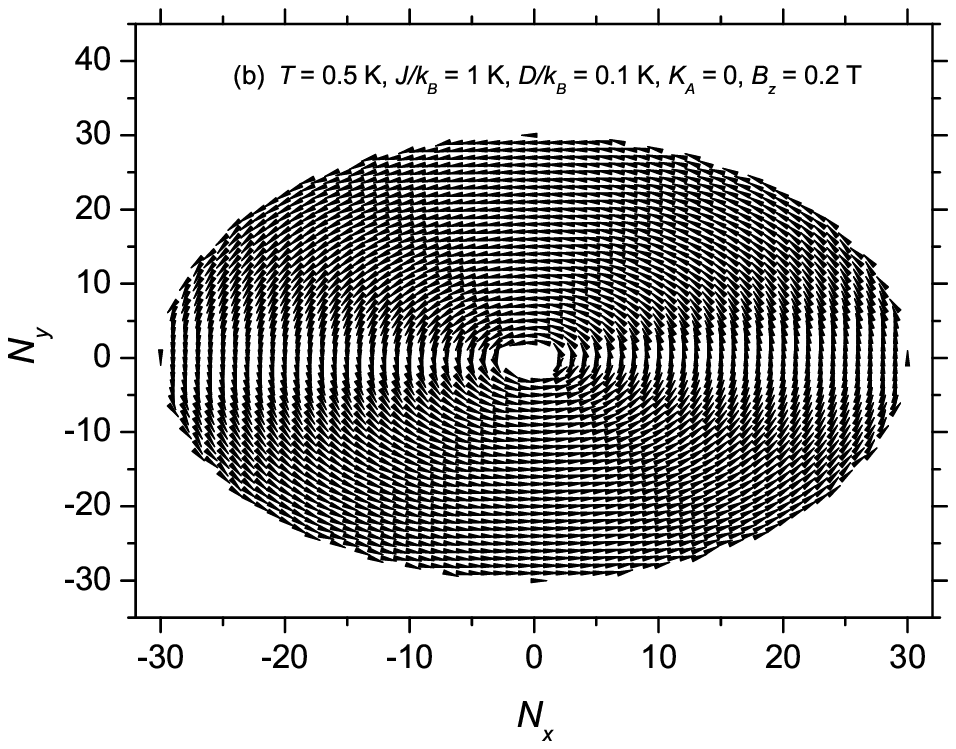,width=.7\textwidth,height=9cm,clip=}
 }
\begin{center}
 \parbox{16cm}
 {\small{{\bf Figure 12.} Calculated spin  configurations projected onto the $xy$-plane
for the nanodisk with $R$ = 30$a$ within a  magnetic field of 0.2
Tesla applied along the $z$-axis, at $T$ = (a)  5 K, and (b) 0.5
K, respectively. Other used parameters are   $D/k_B$ = 0.1 K,
${\cal J}/k_B$ = 1 K, and $K_A$ = 0 K.} }
 \end{center}
\end{figure*}

\begin{figure*}[htb]
\centerline{
 \epsfig{file=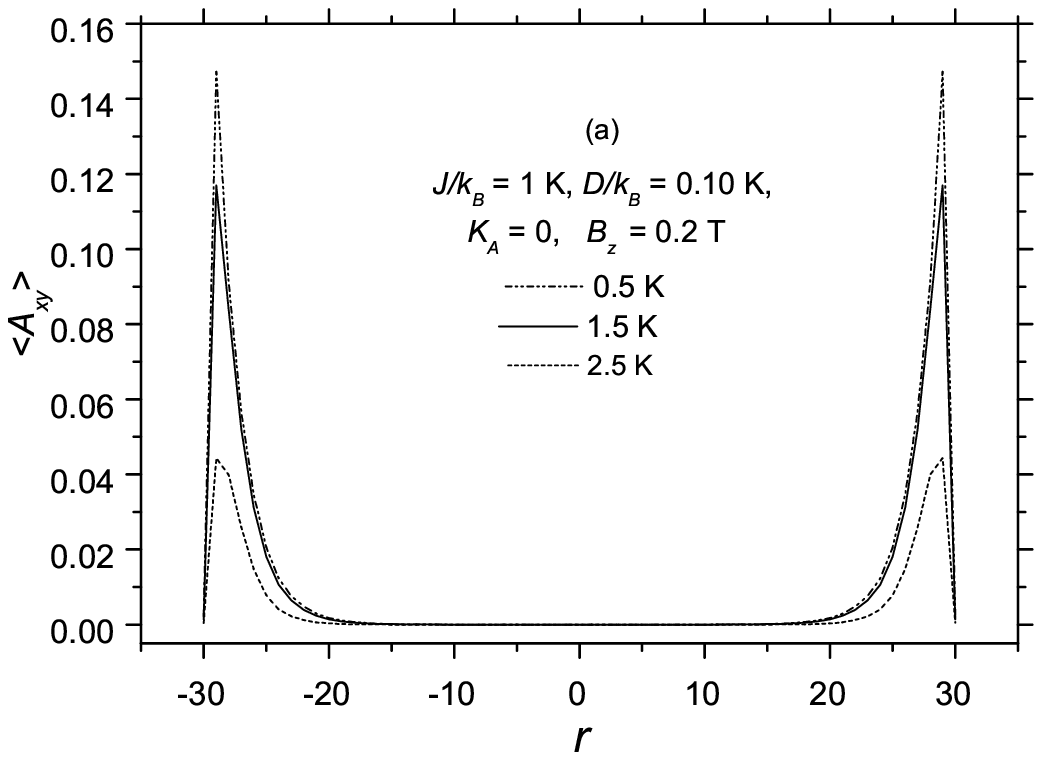,width=.45\textwidth,height=7cm,clip=}
\epsfig{file=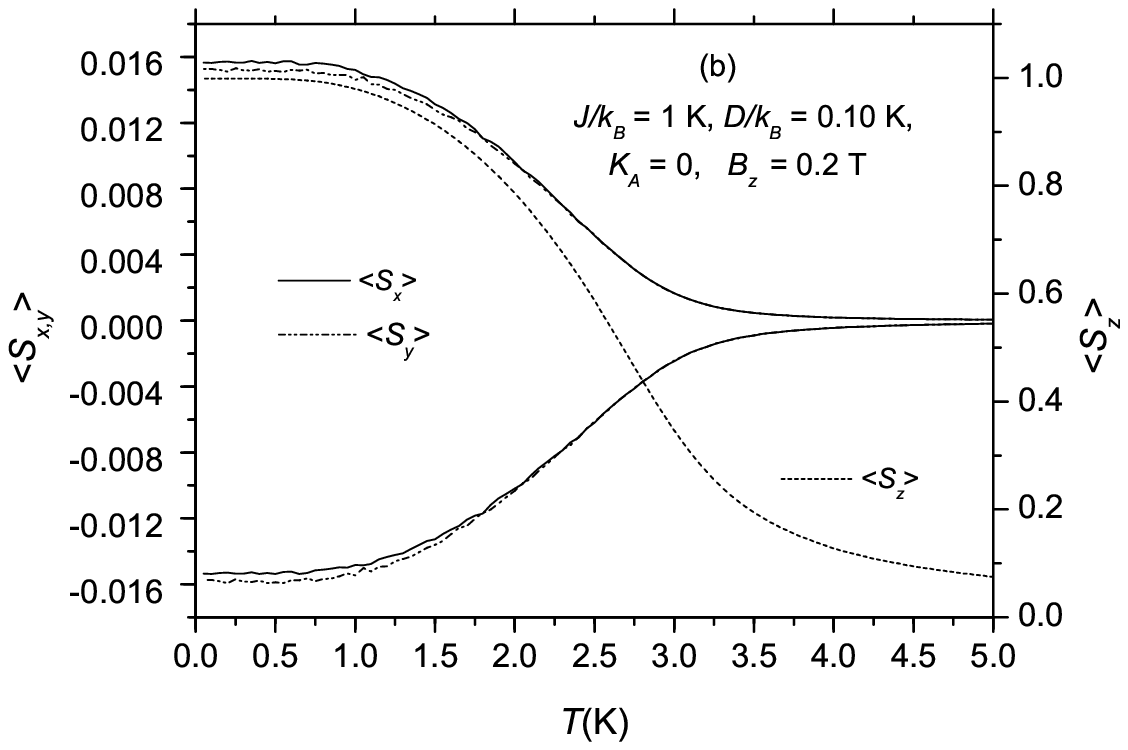,width=.45\textwidth,height=7cm,clip=}}
\begin{center}
 \parbox{16cm}
 {\small{{\bf Figure 13.}  Calculated   (a)    $\langle A_{xy}\rangle $  at different
       temperatures as the functions of $r$,
      (b)   $\langle S_x\rangle$,  $\langle S_y\rangle$ and  $\langle S_z\rangle$
      as the functions of $T$,  assumed in an  magnetic field  of 0.2
       Tesla  applied in the  $z$ direction for the  nanodisk with $R$ = 30$a$.
 Other used parameters are  ${\cal J}/k_B$ = 1 K,  $D/k_B$ = 0.1 K, and $K_A$ = 0.}}
 \end{center}
\end{figure*}

Once again, when an external magnetic field of 0.2 Tesla was
considered to be applied in the $z$-direction to perform
simulations, the two vortices and the strip between them shown in
Figure 11 were merged to a single vortex around the center of the
round disk as depicted in Figure 12. The ring structures there
means that within the blank area the spins have  all been
polarized completely by the applied magnetic field to the
$z$-direction. The Heisenberg and DM interactions compete with the
external magnetic field, and they becomes stronger with decreasing
temperature, so that the width of the ring grows and the blank
area shrinks as $T$ goes down.

Now we define a new quantity as $\langle A_{xy}\rangle =
\overline{\sqrt{\langle S_x\rangle^2+\langle S_y\rangle^2}}$, here
the average is made over all spins of a circle  with radius $r$
around the center. $\langle A_{xy}\rangle$ was calculated at
different temperatures for the circles of different radii, three
of them are now displayed in Figure 13(a). As temperature
decreases, $\langle A_{xy}\rangle$ gets larger, and the ring width
increases. Moreover, at every temperature, $\langle A_{xy}\rangle$
has a maximum near the edge, then fades quickly with decreasing
$|r|$. However, compared to the large magnitude of $\langle
S_z\rangle$ as shown in Figure 13(b), the vortices are actually
very weak.

\subsection{Computational Efficiency}
As described previously, our simulation approach is based on the
principle of the lowest  (free) energy,   so that  a code
implemented with the algorithm is able to minimize the total
(free) energy of the studied system automatically,  and  quickly
converges to the equilibrium state. To test this hypothesis, both
the total energies and total free energies of the   small
 nanodisk with radius $R$ = 10$a$ were recorded during computation at ten temperatures
below 2.8 K as $B_z = 0$, and below 3.8 K when $B_z = $ -0.1 K,
respectively. Consequently, they were all found to decrease
spontaneously and very  quickly toward the equilibrium states
during simulations, so that the computational speed has been
greatly accelerated.  For instance, in the absence of external
magnetic field, as $\tau_0 = 10^{-5}$, $\Delta T$ = -0.02 K,  and
other parameters assigned to the values given below Figure 1, it
only took a few iterations to converge at very low temperatures,
and less than 27 seconds to complete the whole simulation below
the transition temperature $T_M$. Moreover, as an external
magnetic field was assumed to be applied antiparallel to $z$-axis,
the code converged very much faster. The reason might be that an
external magnetic field   is able to effectively suppress the
energy barriers, so that the   computational  code takes much less
time to tunnel through the smoothen energy barriers.

\section{Conclusions and Discussion}

Based on our calculated  results presented above, we can conclude:
(a) The chirality of the single magnetic vortex on the nanodisk
with the co-existence of DM and uniaxial anisotropic interactions
is solely determined by  the sign of $D$: the chirality  is
right-handed as $D > 0$, but left-handed as $D < 0$; (b)  the
external magnetic field applied perpendicular to the $z$-axis can
not change the chirality of the magnetic vortiex, it can only
enhance the magnetizations in the field direction, and increase
the transition temperature so that the vortex structure
  persists to   higher temperatures as observed in
experiments \cite{Yu}; (c)  Increasing the disk size or   DM
interaction strength gives rise to formation  of magnetic domains,
such as magnetic  strips and vortices, on the disk plane, however
  applying  external magnetic field normal
to the disk plane is able to stabilize the vortex structure and
induce skyrmions  \cite{Yu}.

We stress finally that in our SCA model  the spins in the
Hamiltonian are treated as quantum operators, and all physical
quantities have been evaluated with quantum theory. From our
simulations done so far  it can be found that the computational
speed has been considerably accelerated by
 this new computational approach,  and the final simulated results
 are all  reasonable physically  \cite{liujpcm,liuphye141,liuphye142,liupssb},
 especially some of them are well consistent with experiment \cite{liupssb}.
 Therefore, we believe that the new approach can be applied to other
sophisticated magnetic systems where various complicated
interactions exist.

\vspace{0.2cm}
 \centerline{\bf Acknowledgements}
 Z.-S. Liu acknowledges   the financial supports by National Natural Science Foundation of China
 under grant No.~11274177. H. Ian is supported by the FDCT of Macau under grant 013/2013/A1,
University of Macau under grants MRG022/IH/2013/FST and
MYRG2014-00052-FST, and National
 Natural Science Foundation of China under Grant No.~11404415.


\begin{thebibliography}{30}



\bibitem{Dzy} I. E. Dzyaloshinsky, J. Phys. Chem. Solids 4,
241  (1958)

\bibitem{Moriya} T. Moriya, Phys. Rev. Lett. 4, 228  (1960).

\bibitem{Tonomura} A. Tonomura, X. Yu, K. Yanagisawa, T. Matsuda,
Y. Onose, N. Kanazawa, H. S. Park, and Y. Tokura, Nano Lett. 12,
1673 (2012).

\bibitem{Yu} X. Z. Yu, N. Kanazawa, Y. Onose, K. Kimoto, W. Z. Zhang, S. Ishiwata,
Y. Matsui, and Y. Tokura, Nat. Mater. 10, 106 (2011).


\bibitem{Wilhelm}  H. Wilhelm, M. Baenitz, M. Schmidt, U. K. R{\" o}ler, A. Leonov,
and A.N. Bogdanov, Phys. Rev. Lett. 107, 127203 (2011).

\bibitem{Yu2} X. Z. Yu, Y. Onose, N. Kanazawa, J. H. Park, J. H. Han,
Y. Matsui, N. Nagaosa, and Y. Tokurab, Nature 465, 901 (2010).


\bibitem{Muhlbauer} S. M{\" u}hlbauer, B. Binz, F. Jonietz, C. Pfleiderer,
A. Rosch, A. Neubauer, R. Georgii, and P. B{\"o}ni, Science 323,
915 (2009).

\bibitem{Heinze}  Stefan Heinze,   Kirsten Von Bergmann,   Matthias
Menzel,  Jens Brede,  Andr{\'e} Kubetzka,   Roland Wiesendanger,
Gustav Bihlmayer,  Stefan Bl{\" u}gel,
  Nature Physics 7 (9), 713    (2011).


\bibitem{Bode}  M. Bode, M. Heide, K. von Bergmann, P. Ferriani, S. Heinze, G.
Bihlmayer, A. Kubetzka, O. Pietzsch. S. Bluegel and R.
Wiesendanger, Nature 447, 190 (2007)

\bibitem{Skyrme} T. H. R. Skyrme, Nucl. Phys. 31, 556 (1962).


\bibitem{Wachowiak} A. Wachowiak, J. Wiebe, M. Bode, O.
  Pietzsch, M. Morgenstern, and R. Wiesendanger, Science 298, 577
  (2002)

\bibitem{Shinjo} T. Shinjo, T. Okuno, R. Hassdorf, K. Shigeto, and T. Ono,
Science 389, 930 (2000)

\bibitem{Choe} S. B. Choe, Y. Acremann, A. Scholl, A. Bauer, A.
Doran, J. St{\"o}hr, and H. A. Padmore, Science 304, 420 (2004).


\bibitem{Im} M. Y. Im, P. Fischer, K. Yamada, T. Sato, sai, Y.
Nakatani, and T. Ono, Nat. Commun. 3, 983 (2012).


\bibitem{Munzer} W. M{\" u}nzer, A. Neubauer, T. Adams, S.  M{\" u}hlbauer,
C. Franz, F. Jonietz, R. Georgii, P. B{\"o}ni, B. Pedersen, M.
Schmidt, A. Rosch, and C. Pfleiderer, Phys. Rev. B 81, 041203
(2010).

 \bibitem{Kiselev} N. S. Kiselev, A. N. Bogdanov, R. Sch{\"a}fer, and U. K.
R{\"o}$\beta$ler, J. Phys. D: Appl. Phys. 44, 392001 (2011).


\bibitem{liujpcm}  Z.-S. Liu,  V. Sechovsk{\' y}, and M. Divi{\v s},
  { J. Phys.: Condens. Matter} {23}, 016002    (2011).

\bibitem{liuphye141}  Z.-S. Liu,  V. Sechovsk{\' y}, and M. Divi{\v s},
 {Physica E}  59,  27   (2014).


\bibitem{liuphye142}  Z.-S. Liu,  V. Sechovsk{\' y}, and M. Divi{\v
s}, Physica E 62,   123   (2014). 

\bibitem{liupssb}  Z.-S. Liu,  V. Sechovsk{\' y}, and M. Divi{\v s},   {Phys. Status Solidi B}
{  249}, 202    (2012).




\bibitem{YMLuo} Y. M. Luo, C. Zhou, C. Won, and Y. Z. Wu, Effect of Dzyaloshinskii Moriya interaction on magnetic vortex,
arXiv:1401.3292v1.


\bibitem{Kwon13} H. Y. Kwon ,   S. P. Kang,  Y. Z. Wu  and  C. Won  {\textit   J.
Appl. Phys. } {  80}, 133911 (2013).



















\end{thebibliography}
\end{document}